\documentclass{sigkddExp}
\usepackage{tabu} 
\usepackage{float}
\usepackage{hyperref}
\usepackage[stable]{footmisc}
\usepackage{bm}

\usepackage{amssymb}
\usepackage{amsthm}

\theoremstyle{definition}
\newtheorem{exmp}{Example}

\begin{document}
%

\title{Macross: Urban Dynamics Modeling based on Metapath Guided Cross-Modal Embedding}
%

\numberofauthors{4}
%


\author{
%
\alignauthor {Yunan Zhang}\\
       \affaddr{The University of Illinois at Urbana-Champaign}
       \email{yunanz2@illinois.edu}
\alignauthor {Heting Gao}\\
       \affaddr{The University of Illinois at Urbana-Champaign}
       \email{hgao@illinois.edu}
\alignauthor {Tarek F. Abdelzaher}\\
       \affaddr{The University of Illinois at Urbana-Champaign}
       \email{zaher@illinois.edu}
       }
\maketitle
\newcommand{\Corpus}{\textbf{\textit{C}}}
\newcommand{\Vocab}{\textbf{\textit{W}}}
\newcommand{\Times}{\textbf{\textit{T}}}
\newcommand{\Locations}{\textbf{\textit{L}}}
\newcommand{\Graph}{\textbf{\textit{G}}}
\newcommand{\Vertices}{\textbf{\textit{V}}}
\newcommand{\Edges}{\textbf{\textit{E}}}
\newcommand{\x}{\textbf{x}}
\newcommand{\X}{\textbf{X}}
\newcommand{\Y}{\textbf{Y}}
\setlength{\parskip}{1em}
\setlength{\headsep}{0.2in}
\begin{abstract}
As the ongoing rapid urbanization takes place with an ever-increasing speed, fully modeling urban dynamics becomes more and more challenging, but also a necessity for socioeconomic development. It is challenging because human activities and constructions are ubiquitous; urban landscape and life content change anywhere and anytime. It's crucial due to the fact that only up-to-date urban dynamics can enables governors to optimize their city planning strategy and help individuals organize their daily lives in a more efficient way. Previous geographic topic model based methods attempts to solve this problem, but suffer from high computational cost and memory consumption, limiting their scalability to city level applications. Also, strong prior assumptions make such models fail to capture certain patterns by nature. \par

To bridge the gap, we propose Macross, a metapath guided embedding approach to jointly model location, time and text information. Given a dataset of geo-tagged social media posts, we extract and aggregate location and time and construct a heterogeneous information network using the aggregated space and time. Metapath2vec based approach is used to construct vector representations for times, locations and frequent words such that co-occurence pairs of nodes are closer in latent space. The vector representations will be used to infer related time, locations or keywords for a user query. Experiments done on enormous datasets show our model can generate comparable, if not better quality query results compared to state of the art models and outperforms some cutting-edge models for activity recovery and classification.
\end{abstract}
\section{Keywords}
Geographic topic; activity mining; spatiotemporal data; hetergeneous information network; metapath; embedding; representation learning.\par

\section{Introduction}
Living in unprecedentedly big cities, urban citizens are now facing great challenges when they try to find desired place that has interesting activities  take place at a favorable time. Consider a student in a strange metropolis like Los Angeles. He or she might be more interested in whether it is in a quiet neighborhood that seldom hold huge disturbing parties. Or, he or she can be more interested in whether there are special cultural events or other activities held near that region. However, with the ever-changing urban dynamics, no real-estate company's website can capture fully what the region is going on, what might the user be interested in. It is difficult, for not only new "immigrants", but also old residents to efficiently and thoroughly get the information they are interested in. Also,  modeling urban activity is crucial for governors to manage their city since their decisions on assigning resources, urban function planning, transportation planning etc, make sense only if they are catching the up-to-date, fine-grained urban situations. However, traditional approaches like human surveys are costly, time consuming but coarse-grained and get out-of-date quickly. \par

However, recent boom in geo-tagged social media(GTSM) $[1]$ may shed light on such challenges. Social media posts are ubiquitous by nature. People tweet, facebook, write reviews on Yelp at anywhere, anytime, recording interesting moments of some events they are involved, recording what happens in their surroundings. Also, with the popularity of GPS-equipped devices, social media data can be tagged with location information that provides even richer information than simple texts and timestamps. In this way, GTSM provides a multi-aspect description of urban dynamics that can easily be accessed at real-time. Another advantage of GTSM is that usually there will be multiple witness of a single event, making it possible to looking into an event from multiple facet, allowing  a more detailed, thorough grasp of what is really going on. In this way, GTSM provides us with an opportunity to perfectly solve our concern.\par

However, mining GTSM for the information we are interested is far from trivial. Firstly, social media are extremely noisy by nature. Existing studies show that $40\%$ of tweets of non-sense bubbles. It is challenging for both humans and machines to filter such large corpus for the required information. Secondly, even we filter out the useful information, it's still hard to jointly model the location, time and text information together. For example, each modality is of different distributions. Usually it is reasonable to assume Gaussian distributions for location, but not for texts, which are of multinomial distributions. Thus, systematic multi-modal modeling is extremely challenging. Thirdly, social media texts are short and unformal, making it difficult to figure out the real meaning of them due to lack of contextual information and language style that full of OOV.\par

There have been existing fruitful research done in this area that attempt to draw geographic topics topics $[2][3][4]$ from GTSM. However, such methods are not adequate to fulfill out goal to systematically model up-to-date urban dynamics and to provide information retrieval depend on user's personal interests. Geographic topic models, either PLSA based or LDA based, has at least two shortcomings in tackling this task. First, these models are not scalable. Just as other probabilistic topic models, geographic topic model takes long time to train when encountering large corpus, which can hardly be accelerated due to inference of latent variable. But to model city-level dynamics, we need to mine massive GTSM data records. Secondly, these geographic topic models usually rely on strong assumption of latent topic distribution, while real word activities does not necessarily take place following certain distributions. The distribution assumptions make them easily miss important information. \par

Another research line is embedding based methods $[5][6][7]$. This kind of methods fall short in treating every component equally and the embedding methods are fixed instead of flexible to changes, thus can be fragile when one of the components is too sensitive. $[6]$ make a cross-modal embedding of time, location and  a bag of keywords to mine activities based on co-occurrence patterns. However, the embedding of time vector equally with the other two modalities make the model often return outliers that are neither semantic related to other results nor the query word people would be interested in.\par

Considering the unscalablity of topic based models, we attempt to construct an embedding based model. Instead of treating each type of information equally, we incorporate metapath to guide the embedding process. If one type of relation is more important at predicting for certain query, we could adjust the metapath so that it more frequently explore that relation. The embedding results would therefore be more specific to given query task. Our model first aggregates time and space points into hotspot clusters and then use the clusters along with important keywords to construct an undirected heterogeneous network that captures co-occurrence between time, space and words. We then select a metapath to guide the embedding of the obtained network so that the resulting embeddings captures useful semantic information in the network. \par

Our contributions can be summarized as follows:
\begin{enumerate}
    \setlength\itemsep{0em}
    \item Design a clustering algorithm to aggregate spatial and temporal data and identify cluster centers as hot spot.
    \item Construct a heterogeneous network to represent space, time and text correlations
    \item Propose a metapath guided embedding approach to capture semantic relations between time clusters, space clusters and words that allows flexible adjustment of embedding for different query tasks. 
    \item A new benchmark consists a million google map place review from New York City and Los Angeles.
\end{enumerate}
Extensive experiment done on a massive geo-tagged twitter dataset shows the effictiveness of our methods. Our quality study shows our model is able to generate comparable, if not better, results compared to the state of the art methods. Several quantitative analysis shows our model outperform baselines.\\
\\
\section{Related Work}
\textbf{Geographic topic discovery} Geographic topic discovery\\$[2][3][4][8][9][10][11]$ are proposed to find the distribution of topics over regions.$[8]$ is an extension of PLSA, which builds a generative probabilistic model where topic is latent variable, while time, location, words and documents are observable, and then predict life cycle of a given topic at a given location or predict spatial distribution of a topic at a given time. Parameters are updated via EM algorithm.  $[2]$ similarly locations are generated from regions following Gaussian distribution while texts are generated from latent topics following multinomial distributions. $[3]$ improve $[2]$ via ease the distribution assumption on location distributions, making their model capable of capturing non-gaussian distributions. $[4]$ extends LDA via adding coordinate variables over Gasussian distributions. Another direction of this kind of methods incorporate user as a variable into their model. $[11]$ follows the philosophy of $[8]$ but make the model user level, namely incorporate specific user into original generative model. $[9]$ also introduce user variable into generation process. Our research share some siliarity to $[2]$and $[3]$ in that we both models city-level dynamics instead of user level. However, our model differs in previous researches. Our model is embedding based instead of topic model based like previous methods, making our model free from distribution assumptions and poor scalability.\\
\\
\textbf{Urban Segmentation}. This line of research tries to segment urban space via communities formed by similar users or coherent urban functions$[13][14][15][16][17]$. $[13]$use LDA to explain trajectory patterns where transition from two regions are viewed as a word in the vocabulary, while regions of coherent functions are viewed as latent topics that generates such words. $[16]$ extract feature from documents generated by users and areas. Then it cluster semantically coherent user communities and regions. $[15]$ aggregate similar regions and communities via calculating semantic similarity and spatial distances. $[16]$ tries take time into account when finding region segmentations. $[17]$ obtain landscape segmentation through GTSM, clustering temporal landscapes. However, our research differs ffrom theirs in that we are open domain, finding fine-grained activities flexibly from different GTSM instead of find coarse-grained clusters.\\
\\
\textbf{Embedding Methods}There has been a handful researches studying embedding methods$[5][6][18][19][20][21][22]$.$[18]$uses a neural network to encode the neighborhood information of words, which can be used as an efficient way to represent words. $[19][20]$ tries to embedding nodes in graphs via skip-gram. $[21$ follows $[20]$ but make it to hetergeneous information networks via specifying node types. $[5][6][22]$ tries to embed hetergeneous event networks. Our proposed method is similar to $[5][6]$ in that we both tries to embed three modalities, time, location and texts together. But our methods differ from theirs in that our embedding are meta-path guided, thus more flexible to different path selections to avoid too sensitive modalities that may disturb model performance. This kind of embedding has some common sense flaws in that many activities don't have time locality, which means they don't necessarily co-occur at a specific time slots. Also, our model is capable to include more modality, like place type, region topics into the network structure, which can further improve the performance.\\
\\
\textbf{Event detection}This is also a related area to our research, especially local event detection$[5][6][22][24][25]$. This kind of research usually tries to find unusal burst in social streams, which they consider as denoting occurrence of special events. $[5][6][22]$are embedding based methods that cluster spatiotemporal and semantic close posts via measuring distance of their embedded vectors and then classify events from candidates in clusters via burstiness. $[22]$ cluster similar hashtags based on popularity, burstness and localness to clasify events. $[25]$compares posts cluster horizontally to posts at the same time but different location to see if the burst of posts is location specific or is globally distributed, and compare vertically to historical records to see if its routine of time specific at that slot. Our research is different from theirs in that we tries to summarize typical activities at different time and locations while event detection is more interested in unusual activities. Though event detection usually test their model on events like NBA or some concert, our model is able to detect such activities since they usually take place at similar locations at similar periods, where there's a clear cross-modal pattern.
\section{Preliminaries}
In this section, we introduce the preliminary concepts. 
\subsection{Data Processing}
Let $\Corpus$ denote the corpus of our tweet data, $\Vocab$ denote the vocabulary of $\Corpus$ and $r$ denote a single record in the corpus. $r\in\Corpus$. For each record $r$, $r=(t_r,l_r,w_r)$ where $t_r$, $l_r$ and $m_r$ denote time, location and text respectively. $t_r$ is the creation time of the record $r$, $l_r$ is the a latitude and longitude pair representing the creation location of $r$ and $m_r$ is the set of key words contained in $r$ where $m_r\subset\Vocab$. \par

Since the data created at a single time and location is sparse, we cluster time and location to reduce data sparsity. Let $\Times=\{T_{c1},T_{c2},...,T_{ck_t}\}$ denote the set of $k_t$ time clusters and $\Locations=\{L_{c1},L_{c2},...,L_{ck_l}\}$ denote the set of $k_l$ location clusters
We then use the extracted records to construct a undirected time-space-word co-occurrence network. \par

Let $\Graph(\Vertices,\Edges,\Y)$ denote this network. $\Vertices$ is the set of vertices which consists of $3$ types of nodes: time cluster, space cluster and keyword. $\Edges$ is the set of edges and an edge $(v,u)\in \Edges$ exists if and only if $u$ and $v$ co-occurs in the same tweet. 

\begin{exmp}
For a tweet $r=(t_r,l_r,m_r)$ where $t_r\in T_r$, $l_r\in L_r$ and $m_r=\{w_{r1},w_{r2}\}$ we would have edges $(T_r,L_r)$, $(T_r,w_{r1})$, $(T_r,w_{r2})$, $(L_r,w_{r1})$, $(L_r,w_{r2})$ and $(w_{r1},w_{r2})$.
\end{exmp}

Let $weight(u,v)$ denote the weight of edge $(u,v).$ The weight is simply the count of co-occurrence of the two nodes. $\Y$ is a cardinal set that denote the node types. Our network, have $3$ types, i.e., $\Y=\{T,L,W\}$ where $T$, $L$ and $W$ denote type of time, location and word respectively.\par

A metapath $[26]$ is defined as a path on the network schema where the nodes on the paths are types $y\in Y$ instead of vertices $v\in V$. A length $k$ metapath $p$ has the form 
\begin{equation*}
    p=y_1-y_2-...-y_k
\end{equation*}
that represent a composite relation consist of types of $y_1,y_2,...y_k$. We use $p[i]$ to denote $i$th type in the path. A path generated under the guidance of metapath can capture semantic meanings from the network. 
\begin{exmp}
Two path we have experimented with are
\begin{enumerate}
    \setlength\itemsep{0 pt}
    \item $W-W-L-W-W$ attempts to find words co-occur in same location. 
    \item $W-W-L-T-L-W-W$ attempts to find words co-occur in same time and in locations link by the time.
\end{enumerate}
\end{exmp}

\subsection{Model Overview}
Geotemporal-tagged tweets contains additional space and time information in addition to their contents. The tweets tweeted by users are therefore correlated not only semantically, but also spatially and temporally. Intuitively, capturing these correlations should enable us to better model people's activities. Since we have $3$ types of data, i.e., $\Times$, $\Locations$ and $\Vocab$, there are in total $6$ types of relations, namely $\Times-\Times$, $\Locations-\Locations$, $\Vocab-\Vocab$, $\Times-\Locations$, $\Times-\Vocab$, $\Locations-\Vocab$. We make a commonsense assumption that events happened in similar space and time stamp are more likely to correlate to each other and use clustering to capture $\Times-\Times$ and $\Locations-\Locations$ correlations, since times and locations are distributed in continuous space, By aggregating times and locations close in space and largely, the number of nodes we need to compute is reduced and at the same time the data sparsity is alleviated. Mean shift algorithm $[27]$ with kernel density estimation is used for spatial and temporal clustering because its good performance on hot spot detection and its no need for the exact number of clusters. \par
After time and location clustering, we construct a heterogeneous information network with time cluster, location cluster and word as node types in order to capture the rest $\Vocab-\Vocab$, $\Times-\Locations$, $\Times-\Vocab$, $\Locations-\Vocab$ correlations. Metapath $[26]$ has been proposed to capture type information in heterogeneous networks and has been widely applied on task such as similarity measure $[26]$, ranking $[28]$, clustering $[29][30]$ and embedding $[31]$. We adopt Metapath2Vec $[31]$, a metapath guided heterogeneous network embedding with selected metapaths to embed our obtained network $\Graph$. The resulting embedding would capture node similarities such that nodes frequently co-occur on the given metapath is has a close vector representation in the embedding space. \par
Given a user input query, which can be a time, a location or a keyword, we will find the cluster it belongs to and its vector representation in the latent space. Then nearest neighbor is found in time, space and word types as our output. \par
The overall algorithm can be summarized as follows.
\begin{enumerate}
    \item Kernel density estimation and mean shift to find spatial and temporal clusters. 
    \item Construct time-space-word heterogeneous information network.
    \item Select a metapath to measure similarities between same type and different types and perform Metapath2Vec to embed the network.
    \item Given a user input query, find its representation in embedding space and output its top-k nearest neighbors in time, space and vocabulary.
\end{enumerate}

\section{Method}
\subsection{Vocabulary Construction}
For a given tweet corpus, we perform only basic word filtering to construct the vocabulary set because the short text nature of tweets are not suitable for effective phrase mining. We count word frequency in all the tweets and preserve the top $k$ frequent words after filtering out a set of stop words. In our experiments, $k$ is set to be $20000$.
\subsection{Spatial and Temporal Clustering}
\begin{figure}[h]
\centering
\includegraphics[width=0.45\textwidth]{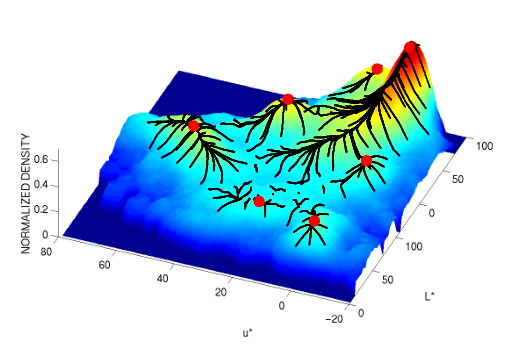}
\caption{Illustration of Mean Shift $[27]$}
\end{figure}
Different for vocabulary, times and locations are distributed in continuous space. Therefore, there are very few data in a single time or space point and these single points usually does not contain distinctive information. In order to reduce data sparsity problem, we aggregate times and locations that are close to each other into clusters and use the aggregated clusters as basic unit for network construction later. \par
We use a kernel function $[27]$ to estimate the density at a given point. Given a set of $n$ data points $\{\textbf{x}_i\}^n_0$ where $\textbf{x}_i\in\mathbb{R}^d$, kernel bandwidth $h$, the kernel density $f(x)$ at an arbitrary point $\textbf{x}$ is given as:
\begin{equation*}
    f(x)=\displaystyle\frac{1}{nh^d}\displaystyle\sum_{i=1}^{n}K(\displaystyle\frac{\textbf{x}_i-\textbf{x}}{h})
\end{equation*}
The $K$ denote the kernel function. There are plenty of kernel function studied in detail and in our settings, but we find in our experiments, the choice of kernel function is not critical to performance. we use Radial basis function kernel $[27]$ given as:
\begin{equation*}
    K(\textbf{x},w)=\frac{C}{w^d}e^{-\frac{||\textbf{x}||_2}{2w2}}
\end{equation*}
where $C$ is normalization constant and $w$ the bandwidth of kernel function. Here we set $w$ to $1$ because we have picked the bandwidth $h$ in density estimation process. \par
We then use mean shift algorithm to find clusters and cluster centers. Mean shift algorithm iteratively shift the center of a cluster to the direction that can maximize the kernel density. 
\begin{exmp}
Given a set of points $\X=\{\x_1,\x_2,...,\x_k\}$ in a window of size $h$, and their current center $\x_c^{(t)}$. Let $\Delta\x$ denote the shift vector. The shift vector is a weighted sum of all the point in the window where the weights are calculated by applying the kernel function on the distance vectors between current center $\x_c^{(t)}$ and the data points. In other words,
\begin{eqnarray*}
    \Delta\x&=&\displaystyle\sum_{i=1}^k w_i \x_i \\
    &=&\displaystyle\frac{1}{C}\displaystyle\sum_{i=1}^k f(\x_i-\x_c) \x_i
\end{eqnarray*}
where $C$ is a normalization constant. \par
The next center $\x_c^{(t+1)}$ will be the current center $\x_c^{(t)}$ shifted by $\Delta\x$, i.e., $\x_c^{(t+1)}=\x_c^{(t)}+\Delta\x$. This process is repeated until convergence.
\end{exmp}

After several iterations, the cluster center would converge to one of the local kernel density maxima. We apply mean shift algorithm on every single data point to find all interesting local density maxima as our cluster centers. \par
This straigthforward approach is not going to scale very well because the computation time is $O(N^2)$. Given a large dataset containing millions of tweets, shifting the mean for each single location is not intractable. An acceleration method $[32]$ is adopted to partition space into grids and each grid contains the number of points and the sum of all point data. This discretization process takes $O(N)$ time and the number of nodes for clustering would be greatly reduced. 
\subsection{Network Embedding}
\subsubsection{Network Construction}
We constructed a heterogeneous information network by treating each time cluster, each location cluster and each word as node and by treating the each co-occurrence relation as edge. The edge weight is the number of times that the two nodes co-occur in the same tweets.
\subsubsection{Metapath Guided Random Walk}
In order to capture the similarity between nodes of the same type and nodes of different types, we perform metapath guided random walks to generated paths. Let $u_y$ denote a node of type $y$ and $\phi(u)$ denote the type of node $u$. $\phi(u_y)=y$ Given a length-$k$ metapath $p$ and a starting node $u^{(0)}_{p[0]}$, the path generated will be $[u^{(0)}_{p[0]},u^{(0)}_{p[0]},...,u^{(k-1)}_{p[k-1]}]$ where the subscript $p[i]$ is the $i$th node type in the metapath. We use $u^{(i)}$ to represent $u^{(i)}_{p[i]}$ for short. $p(u^{(i+1)}|u^{(i)})$ denotes the transition probability of going from node $u^{(i)}$ to neighbor $u^{(i+1)}$. $p(u^{(i+1)}|u^{(i)})$ is proportional to the edge weight of the two nodes. The formal definition is as follows.
\begin{equation}
    p(u^{(i+1)}|u^{(i)})=\begin{cases}
                        \frac{1}{C}weight(u^{(i)},u^{(i+1)}) & if \phi(u^{(i+1)})=p[i+1]\\
                        0\mbox{, if }(u^{(i)},u^{(i+1)})\notin\Edges\\
                        0\mbox{, if }u^{(i+1)}\neq p[i+1]
                        \end{cases}
\end{equation}
where $C$ is a normalization factor.
\begin{exmp}
For a given metapath $p=W-W-L-W-W$, we find all the nodes of the first type, which is $W$, in the metapath. In this case, we would find all the words $w\in\Vocab$ as our starting nodes. We then move to check the second type, which is also $W$. For each word $w$ in the starting nodes, we find all its neighbor nodes $N={w_1,w_2,...,w_k}$ of the second type $W$. We would then go to $w_i$ with a probability proportional to the edge weight of $(w,w_i)$. After we get the second node, we check the third type in the metapath and so on. This process is repeated until we finish traversing the metapath. 
\end{exmp}

\subsubsection{Metapath Guided Embedding}
By going through rounds of random walks with a given metapath, we obtains a set of random paths. These random path contains semantic information of the metapath and can be used as skip-grams to embed the whole heterogeneous network. We use an embedding method similar to Metapath2Vec [2] to embed our graph, which is the state of art metapath guided node embedding algorithm for heterogeneous information network. It takes a set of metapath guided random path as input and output a latent vector representation for each nodes. Each path is treated as a short document and skip-gram model and continuous bag of words model is used to capture correlations between nodes. \par
heterogeneous negative sampling strategy is used to improve accuracy on heterogeneous embeddings and computational efficiency. Previous methods such as Deepwalk [] and node2vec [] embeds nodes in homogeneous graphs and does not need to consider type information. However, things becomes different once we want to embed a heterogeneous network. The probability calculation for node content should be specific to the node type and be independent of other types of nodes. Similar to Metapath2Vec, we use the modified version of the previous node probability definition as 
\begin{eqnarray*}
    p(c_t|v)=\frac{e^{X_{c_t}X_{v}}}{\sum_{u_t\in\Vertices_t}e^{X_{u_t}X_{v}}}
\end{eqnarray*}
where $c_t$ is the content or the neighbor node of the given node $v$ of a given node type $t$. $X_v$ is vector representation of a node $v$. $\Vertices_t$ is all the vertices of node type $t$. The probability of a content node is therefore normalized against the node of the same type instead of against all the nodes in the graph. \par
Although normalization within types can reduce the computational complexity, the probability model is still intractable to compute. An ordinary approach is to use negative sampling to approximate the probability. Negative sampling randomly samples $M$ nodes that are not neighbors of the given node $v$ and approximate the objective function as follows
\begin{eqnarray*}
    O=log\sigma(X_{c_t}\cdot X_v)+\displaystyle\sum_{m=1}^M\sim p(u_t^m) log\sigma(-X_{u_t^m}\cdot X_v)
\end{eqnarray*}
The gradient of the objection function is easy to compute and stochastic gradient descent is used to optimize the model. \par
\subsection{Output Related Node}
The embeddings of our time-location-word network can then be used to perform prediction task since semantically related nodes are close in latent space. Given a user input query, we find the cluster it belongs to if the query is a time or location or find the word in the graph. Then nearest neighbor algorithm is performed to get the most related times, locations, and words as the output of the query.

\section{Experiments}
\subsection{Experiment Setup and Datasets}
\textbf{Data Sets.} We run our experiments on two enormous real word datasets: 1) Geotagged Tweet Dataset collect from Twitter, which consists of around 2.3 million geo-tagged tweets posted in Los Angeles during 2014.08.01 \- 2014.11.30 and 2015.08.01 \- 2015.11.30. 2) Google Map Place Review Dataset. We crawled this dataset with our crawler that can crawl all the google map place reviews in a given radius. It consists of around 1 million reviews from New York City, which include location coordinates, timestamps, review contents, user rating of the place and the place type. We clear infrequent words that occur less than 100 times and stopwords from both datasets.\\

\textbf{Baselines} We compare Macross with the following methods.

\begin{itemize}
\item LGTA$[2]$ is a PLSA based geographic model that generate location over latent regions following Gaussian and words over latent topics following multinomial.\\
\item MGTM$[3]$ is a SotA geographic topic model. It adopt multi-Dirichlet process to enable the model to find non-gaussian distributed that $[2]$cannot find.\\
\item CrossMap$[6]$ is a SotA multi-modal embedding approach that tries to model urban dynamics via cooccurrence patterns of location, time and texts.\\
\end{itemize}
\textbf{Parameter Setting}
We set location bandwith as 0.05, time bandwidth as 1000, vocabulary size as 20000, embedding vector size as 300 dimension, negative sample as 5, window size as 7, walk length as 50, the number walk as 30. 
\subsection{Results}
\subsubsection{Quality Study and Illustrative Cases}

We first test our model on the Twitter dataset to see if it can really capture the co-occurrence patterns of time, location and keywords. When query some activity names, our model will return the usual locations and times these activities take place as well as topical phrases that can represent the entered activities. We compare our results to those returned by CrossMap, analysis the quality and resonability of the return to see if there's improvement.

\begin{figure}[h]
\centering
\includegraphics[width=0.5\textwidth]{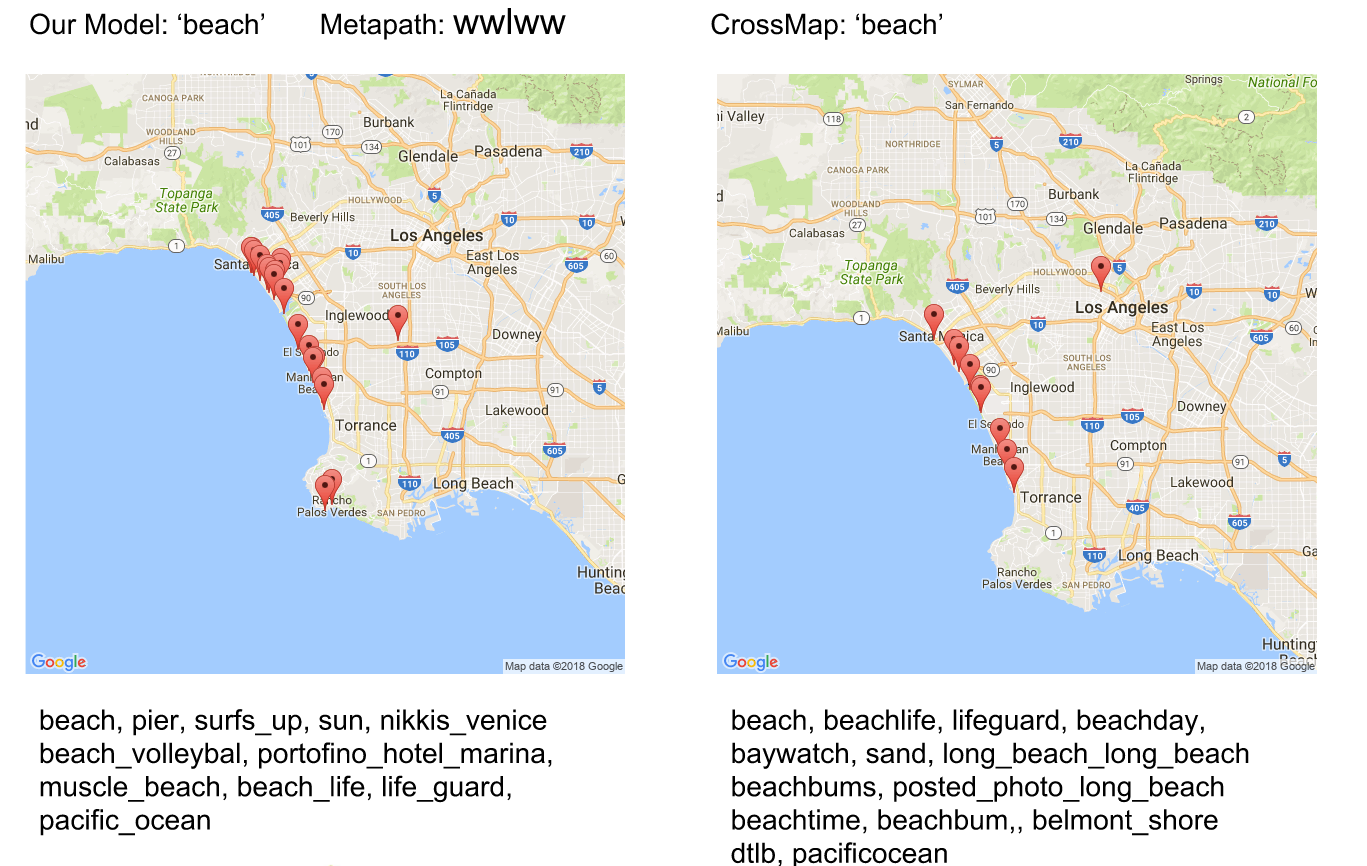}
\caption{query "beach" comparison between Macross and CrossMap}
\end{figure}

\begin{figure}[h]
\centering
\includegraphics[width=0.5\textwidth]{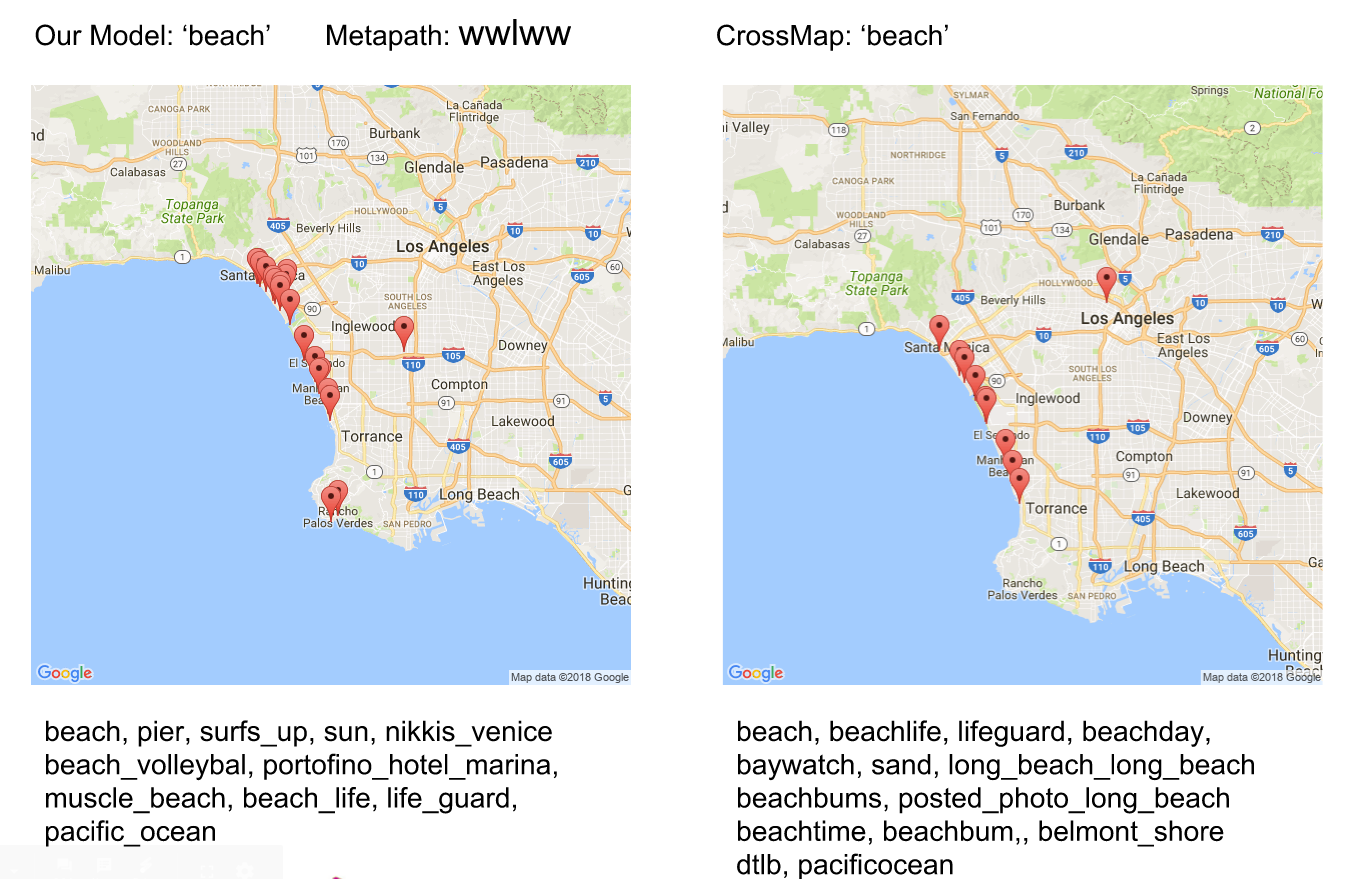}
\caption{query "food" comparison between Macross and CrossMap}
\end{figure}

\begin{figure}[h]
\centering
\includegraphics[width=0.5\textwidth]{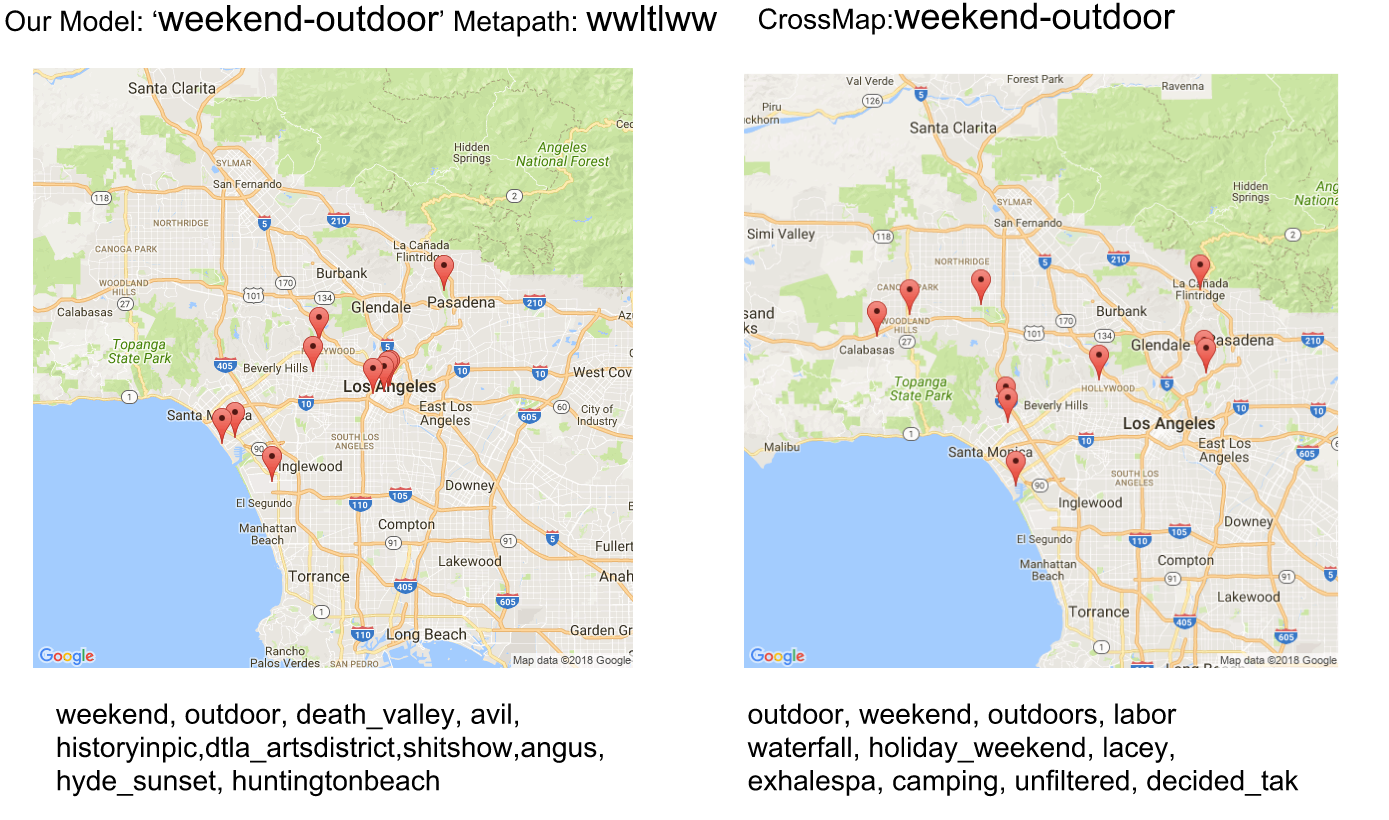}
\caption{query "weekend" comparison between Macross and CrossMap}
\end{figure}

\begin{figure}[h]
\centering
\includegraphics[width=0.5\textwidth]{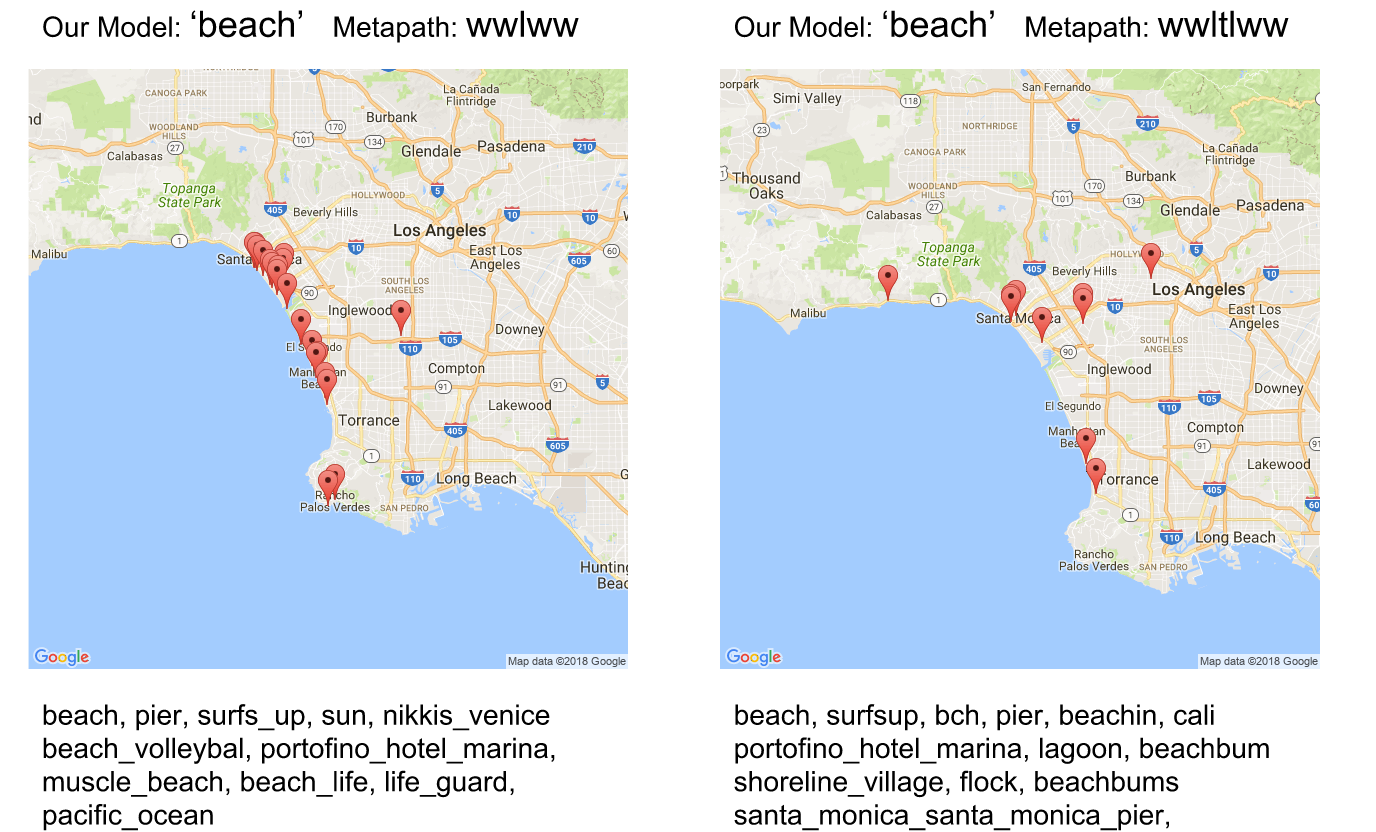}
\caption{query "beach" comparison between different metapath used in Macross}
\end{figure}

Figure 2 shows the result when we query 'beach'. As we can observe above,  Macross find almost all the beach locations on the west coastline and several points on the south where there's famous 'Long Beach'.  However, CrossMap find few beaches on the west coastline and totally misses the Long Beach. Also, when we looking into topical phrases returned by our model, we see that CrossMap tends to return synonyms  of the query word such as beach, beachlife, beachday while our model return many geo-specific words like nikks venice, beach volleyball, portofina hotel marina and muscle beach. \\
\\
Figure 3 showsthe result when we query 'food'. As we checked on google map, there are popular restaurants in locations returned by both models. But still we can see the topical phrases returned by Macross, is comparable, if not better than those returned by CrossMap.\\
\\
Figure 4 shows the result when we enter 'weekend-outdoor'. We can see that in this one, CrossMap returns some irrelevant words like labor, lacey and unfiltered. On contrast, our model returns high quality phrases like death valley, dtla art district, hyde sunset that are special outdoor attractions in Los Angeles.
\subsubsection{Quantitative Evaluation}
We use mean reciprocal rank (MRR) as our measure of performance, which is defined as follows
\begin{equation*}
    MRR=\frac{1}{|Q|}\displaystyle\sum_i^{|Q|}\frac{1}{r_i} \\
\end{equation*}
where $Q$ is a set of queries as ground truth and $r_i$ is the output rank of $i$th query(ground truth). The higher the mean reciprocal rank, the better the performance.
\begin{exmp}
We use a test set of tweets data that contains tuples $(T,L,W)$ pairs as ground truth. For a given query, say a word "beach", we first use our model to predict the related keywords and locations as output results. We then find in the test data the true related words and locations as ground truth. The reciprocal rank is calculated for the query "beach". Same procedural is done for other queries. We then take average of all the scores.
\end{exmp}

\begin{figure}[h]
\centering
\includegraphics[width=0.5\textwidth]{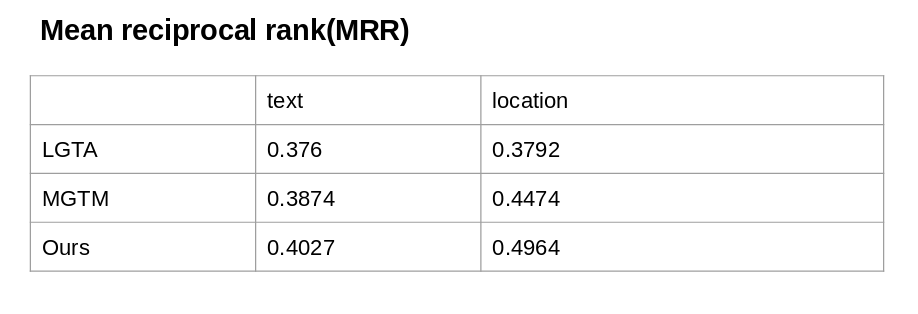}
\caption{Mean Reciprocal Rank Comparison on Text and Location}
\end{figure}

\section{Conclusion and Future Work}
We have studied the problem of modeling urban dynamics using GTSM. We propose Macross, which is a meta-path guided multi-modal embedding method that can provides an efficient and effective representation of urban dynamics. Time, location and texts jointly embedded into the same space to learn their co-occurrence patterns while embedding path selection is flexible enough to avoid too sensitive variables. As a result, Macross can not only fully capture rapid urban changes, but can also bring about lots of downstreaming applications.

For feture work, we consider adopt reinforcement learning to automatically find the best path, since currently, we observe different meta-path selection seems to have different performance according to the query type, which requires human tuning to get optimized. Also, we choose meta-path guide embedding in this workm but motif may has richer information than meta-path, we may experiment motif-based methods instead of meta-path based in the future. 
\section{Acknowledgements}
This work receives help from Chao Zhang from Data Mining Group.

\section{Reference}
$[1]$C. Zhang, Q. Yuan and J. Han, "Bringing Semantics to Spatiotemporal Data Mining: Challenges, Methods, and Applications," 2017 IEEE 33rd International Conference on Data Engineering (ICDE), San Diego, CA, 2017, pp. 1455-1458.doi: 10.1109/ICDE.2017.210\\
$[2]$Z. Yin, L. Cao, J. Han, C. Zhai, and T. Huang. Geographical topic discovery and comparison. In Proceedings of WWW 2011, pages 247–256, New York,\\
$[3]$ C. C. Kling, J. Kunegis, S. Sizov, and S. Staab. Detecting
non-gaussian geographical topics in tagged photo\\
$[4]$S. Sizov. Geofolk:  latent spatial semantics in web 2.0 social media. In
WSDM, pages 281–290, 2010.
$[5]$Zhang, C.; Zhang, K.; Yuan, Q.; Tao, F.; Zhang, L.; Hanratty, T.; and Han, J. 2017c. React: Online multimodal embedding for recency-aware spatiotemporal activity modeling.  In SIGIR , 245–254.\\
$[6]$C.  Zhang,  K.  Zhang,  Q.  Yuan,  H.  Peng,  Y.  Zheng,  T.  Hanratty,
S.  Wang,  and  J.  Han,  “Regions,  periods,  activities:  Uncovering
urban  dynamics  via  cross-modal  representation  learning,”  in
WWW, 2017, pp. 361–370\\
$[7]$Zhang, C.; Liu, L.; Lei, D.; Yuan, Q.; Zhuang, H.; Hanratty,
T.; and Han, J.  2017a.  Triovecevent: Embedding-based on-
line local event detection in geo-tagged tweet streams.   In KDD, 595–604.\\
$[8]$Q.  Mei,  C.  Liu,  H.  Su,  and  C.  Zhai.A  probabilistic  approach  to spatiotemporal theme pattern mining on weblogs. In WWW, pages 533–542, 2006\\
$[9]$Q. Yuan, G. Cong, Z. Ma, A. Sun, and N. M. Thalmann. Who, where,when and what: discover spatio-temporal topics for twitter users. In KDD , pages 605–613, 2013\\
$[10]$ P. Wang, P. Zhang, C. Zhou, Z. Li, and G. Li.  Modeling infinite topics on  social  behavior  data  with  spatio-temporal  dependence.   In
CIKM, pages 1919–1922. ACM, 2015.\\
$[11]$Kurashima, T.; Hoshide, T.; Takaya, N.; Fujimura, K.; Iwata, T.Geo topic model: joint modeling of user's activity area and interests for location recommendation. WSDM 2013 - Proceedings of the 6th ACM International Conference on Web Search and Data Mining 2013, pp. 375-
Scopus, Conference.\\
$[12]$ L. Hong, A. Ahmed, S. Gurumurthy, A. J. Smola, and
K. Tsioutsiouliklis. Discovering geographical topics in the
twitter stream. In WWW , pages 769–778, 2012\\
$[13]$J.  Yuan,  Y.  Zheng,  and  X.  Xie.Discovering  regions  of  different functions  in  a  city  using  human  mobility  and  pois.   In KDD,  pages 186–194, 2012.\\
$[14]$ V. Frias-Martinez, V. Soto, H. Hohwald, and E. Frias-Martinez. Characterizing urban landscapes using
geolocated tweets. In SocialCom/PASSAT, pages 239–248,2012.\\
$[15]$J. Cranshaw, R. Schwartz, J. I. Hong, and N. Sadeh. The livehoods project:  Utilizing social media to understand the dynamics of a city. In ICWSM, pages 58 – 65, 2012.\\
$[16]$A. Noulas, S. Scellato, C. Mascolo, and M. Pontil.Exploiting semantic annotations for clustering geographic areas and users in location-based social networks. In ICWSM, 2011.\\
$[17]$  K. Zhang, Q. Jin, K. Pelechrinis, and T. Lappas. On the importance of temporal dynamics in modeling urban activity. In UrbComp, 2013.\\
$[18]$T. Mikolov, I. Sutskever, K. Chen, G. Corrado, and J. Dean. Distributed Representations of Words and Phrases and their Compositionality. Accepted to NIPS 2013.\\
$[19]$Aditya Grover and Jure Leskovec. 2016. node2vec: Scalable feature learning for networks. In KDD. ACM, 855–864.\\
$[20]$ J. Tang, M. Qu, M. Wang, M. Zhang, J. Yan, and Q. Mei. Line:  Large-scale information network embedding. In WWW, pages 1067–1077, 2015.
$[21]$Y. Dong, N. V. Chawla, and A. Swami, “metapath2vec: Scalable representation  learning  for  heterogeneous  networks,”  in KDD,2017, pp. 135–144.
$[22]$ H. Gui, J. Liu, F. Tao, M. Jiang, B. Norick, and J. Han.Large-scale embedding learning in heterogeneous event data. In
ICDM, 2016.\\
$[23]$R. Lee, S. Wakamiya, and K. Sumiya. Discovery of unusual regional social activities using geo-tagged microblogs. World Wide Web
, 14(4):321–349, 2011\\
$[24]$ W. Feng, C. Zhang, W. Zhang, J. Han, J. Wang,C. Aggarwal, and J. Huang. Streamcube:  hierarchical spatio-temporal hashtag clustering for event exploration over the twitter stream. In ICDE , pages 1561–1572, 2015.\\
$[25]$C. Zhang, G. Zhou, Q. Yuan, H. Zhuang, Y. Zheng, L. Kaplan, S. Wang, and J. Han.Geoburst: Real-time local event detection in geo-tagged tweet streams. In SIGIR, pages 513–522, 2016\\
$[26]$Y. Sun, J. Han, X. Yan, P. S. Yu, and T. Wu.  Pathsim: Meta path-based top-k similarity search in heterogeneous information networks. In VLDB’ 11, 2011. \\
$[27]$D. Comaniciu and P. Meer. Mean shift:  A robust approach toward feature space analysis. IEEE Trans. Pattern Anal.Mach. Intell., 24(5):603–619, 2002. \\ 
$[28]$Y. Sun, J. Han, P. Zhao, Z. Yin, H. Cheng, and T. Wu. Rankclus: Integrating clustering with ranking for heterogenous information network analysis. In Proc. 2009 Int. Conf. Extending Database Technology (EDBT’09) , pages 565–576, Saint Petersburg, Russia, Mar. 2009. \\
$[29]$Yizhou Sun, Brandon Norick, Jiawei Han, Xifeng Yan, Philip S. Yu, and Xiao Yu, "Integrating Meta-Path Selection with User Guided Object Clustering in Heterogeneous Information Networks", Proc. of 2012 ACM SIGKDD Int. Conf. on Knowledge Discovery and Data Mining (KDD'12), Beijing, China, Aug. 2012\\
$[30]$Ming Ji, Jiawei Han, and Marina Danilevsky, "Ranking-Based Classification of Heterogeneous Information Networks", Proc. of 2011 ACM SIGKDD Int. Conf. on Knowledge Discovery and Data Mining (KDD'11), San Diego, Aug. 2011\\
$[31]$Dong, N. V. Chawla, and A. Swami, “metapath2vec: Scalable representation learning for heterogeneous networks,” in KDD,2017, pp.135-144.\\
$[32]$M. A. Carreira-Perpinan. Acceleration strategies for gaussian mean-shift image segmentation. In CVPR, pages 1160–1167, 2006. \\

\end{document}